\documentclass{article}
\usepackage[utf8]{inputenc}

\usepackage[
    backend=bibtex,
    style=authoryear,
    natbib=true,
]{biblatex}

\makeatletter
\newcommand*{\centerfloat}{%
  \parindent \z@
  \leftskip \z@ \@plus 1fil \@minus \textwidth
  \rightskip\leftskip
  \parfillskip \z@skip}
\makeatother

\usepackage[colorlinks=true, allcolors=blue]{hyperref}
\usepackage{adjustbox}

\usepackage{amsmath}
\usepackage{graphicx}
\usepackage{float}
\usepackage{geometry}
\usepackage{authblk}
\usepackage{array}
\usepackage{aas_macros}
\usepackage{rotating}

\title{Non-spherical dark matter structures detection}

\date{\today}

\author{Nicolas Loizeau}
\author{Glennys R. Farrar}
\affil{\small\it{Center for Cosmology and Particle Physics, New York University}}

\begin{document}

\maketitle

\begin{abstract}
    A rotation curve inequality that holds for spherically symmetric mass distributions is derived, and tested against the SPARC galaxy rotation curves dataset. We identify several Galaxies, eg NGC7793 and UGC11557, which are candidates for hosting non-spherical dark matter structures that could be detected by more precise measurements.
\end{abstract}

\section{Introduction}
\label{inequalities}
Cosmological simulations predict that the dark matter halos of low-redshift galaxies are triaxial on large scales.  In addition to such very large scale asphericity, an interaction between dark matter and baryons could lead to an accumulation of dark matter in regions of gas overdensities, resulting in a (puffy) DM disk or other  structure in the inner galaxy.  Here we propose a strategy for detecting such small-scale DM structures if they exist.

Energy transfer is most efficient in collisions between similar mass particles, so a DM accumulation in the vicinity of baryons would be greatest for DM whose mass is of order the mass of H and He, and likely undetectably small for significantly lower and higher mass DM. The random velocities of DM particles in the halo are generally larger than those of gas in the disk, so on average a DM-baryon collision causes the DM particle final phase space to approach that of the initial baryon.  The scattering heats the gas, which then re-cools through standard mechanisms. (An almost infinitesimal fraction of DM particles intersect a star; those would just reflect or more usually be trapped in the star, negligibly impacting the global DM distribution.)  The net DM accumulation would be greatest around the longest-lived gas structures and of course would be proportional to the cross-section between DM and baryons.

Our purpose here is to introduce a practical and unambiguous diagnostic of structure in the distribution of DM on small scales, as could result from DM-baryon interactions.

Rotation curves are central tools to study the mass distribution of individual galaxies \citep{Rubin1980}.  The dark matter distribution can be estimated by subtracting the baryonic components from the observed mass distribution.
There is a degeneracy in the mass distribution determination using rotation curves: two different mass distributions can lead to the same rotation curve. This degeneracy can be broken by imposing that the DM distribution have a particular functional form. Common DM halo models are spherically symmetric. A given DM distribution model can be favored/excluded because it gives better/worse rotation curves fits, but being able to rigorously exclude a model based on incompatibilities with some rotation curve features would be a powerful tool. 
One feature to consider is sphericity.
Whether DM disks, or  DM structures such as non-monotonic radial density exist, is still an open question \citep{Fan2013, Schutz2018, Buch2019, McCullough2013}.

In a previous paper, we fit 121 galaxy rotation curves with a hadronically interacting dark matter (HIDM) model composed of a DM halo and a DM disk that scales the halo-baryons interactions \citep{Loizeau2021}. We showed that this model outperforms 7 other canonical DM and modified gravity models. This finding motivates the need for having other, model independent, indicators of non-spherical dark mater structures in galaxies.  (Note that the analysis does not reveal the thickness of the DM disk, which is presumably thick.)

Some non-spherical mass distributions can produce rotation curves that are impossible to build from a spherical distribution. For example, a ring of mass produces an outward gravitational pull inside of the ring, a phenomena that is impossible to reproduce with a spherical mass distribution. 
We propose a method to detect some such asphericities and test it on the SPARC rotation curves dataset \citep{Lelli2016}.

In sections \ref{sec:inequalities} and \ref{examples} we derive a general asphericity diagnostic for rotation curves and test it on some ideal mass distributions. In section \ref{sec:sparc} we evaluate the asphericity diagnostic for the SPARC galaxy rotation curves and show how the asphericity diagnostic is related to our previously developed HIDM model. The diagnostic is defined independently of any specific mass model. 

\section{Spherical inequalities}
\label{sec:inequalities}
We derive spherical inequalities that can be violated when the rotation curve admits some asphericities. These inequalities are only based on the rotation curve $v(r)$ and are model-independent. This makes them the perfect tool for asphericity detection in well measured rotation curves. However not all asphericities violate the inequality. Thus, if the inequality is violated (at an observationally robust level), asphericity of the DM halo is rigorously established, but the halo can be aspherical without violating the inequality.
\subsection*{Continuous}
Consider a spherical mass distribution : $v=\sqrt{\frac{GM}{r}}$ where $M$ is the total mass inside the sphere of radius $r$. Then, $\frac{dv}{dr}=\frac{\sqrt{G}}{2}\left(\frac{1}{r}\frac{dM}{dr}-\frac{M}{r^2}\right)\sqrt{\frac{r}{M}}$, so
\begin{equation}
    \frac{dv}{dr} = \frac{1}{2}\sqrt{\frac{G}{rM}}\frac{dM}{dr}-\frac{v}{2r}.
\end{equation}
Hence, since $\frac{dM}{dr}\geq 0$, we have the inequality

\begin{align}
    \Delta(r)&\equiv1+\frac{v}{2r}\left(\frac{dv}{dr}\right)^{-1} \geq 0 \text{, i.e.,}\\
    &=1+\frac{d\ln r}{d\ln v} \geq 0
    \label{delta}
\end{align} 
at all radii for any spherical mass distribution.
Note $\Delta<0$ implies that the mass distribution is non-spherical but the existence of a non-spherical mass distribution does not imply $\Delta<0$. For example, the Mestel disk \citep{Mestel1963, Binney1987} is a disk distribution that has the same rotation curve as a point mass and therefore does not violate the spherical inequality.

\subsection*{Discrete}
It is usefully to write the spherical inequality for discrete measurements since observations are at a discrete set of radii. Let $i$ and $j$ be 2 data points of a rotation curve with $r_j>r_i$. Then we have $M_i \leq M_j$ where $M_{i,j}$ is the total mass inside the sphere of radius $r_{i,j}$. Assume that the mass distribution is spherical : $M_{i,j}=v_{i,j}^2 r_{i,j} /G$. This leads to the inequality $v_i^2 r_i \leq v_j^2 r_j $ and 
\begin{equation}
    \Delta_{ij} \equiv 1-\frac{v_i}{v_j}\sqrt{\frac{r_i}{r_j}} \geq 0
    \label{delta_discrete}
\end{equation}
 for any spherical mass distribution. Asphericity diagnostic $\Delta$ can be used to detect asphericity: 
negative $\Delta_{ij}$ implies that there exists some non-sperical structure such as a disk or a ring. Let $\delta v$ be the error on velocity $v$. Then, the error on $\Delta_{ij}$ is
\begin{equation}
    \delta\Delta_{ij} = \sqrt{\frac{1}{v_j^2}\frac{r_i}{r_j}(\delta v_i)^2+\frac{v_i^2}{v_j^4}\frac{r_i}{r_j}(\delta v_j)^2}.
    \label{eq:discrete_error}
\end{equation}

\section{Examples}
\label{sec:examples}
\label{examples}
\begin{figure}[H]
\centering
\includegraphics[width=1\textwidth]{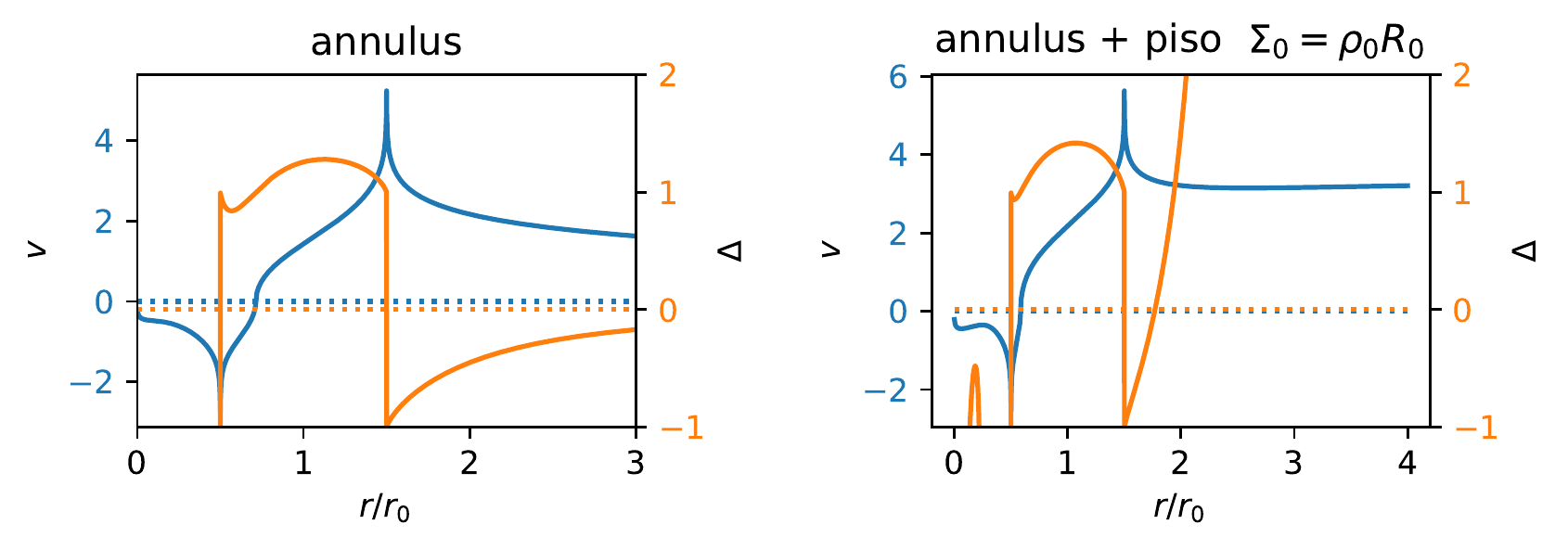}
\caption{\textit{Left}: blue: rotation curve of an annulus of inner radius $r_0/2$ and outer radius $3r_0/2$. Orange: corresponding asphericity diagnostic $\Delta$. The inner negative rotation curve means that a test particle is pulled outward. This phenomena occurs in gas contributions to particular SPARC rotation curves when there are gas over-densities at a certain radius. \textit{Right}: rotation curve and asphericity diagnostic for the superposition of a piso halo and an annulus with the same scale radius $r_0$, width $d=r_0$ and $\Sigma_0=\rho_0r_0$.}
\label{fig:annulus}
\end{figure}

\begin{figure}[H]
\centering
\includegraphics[width=0.8\textwidth]{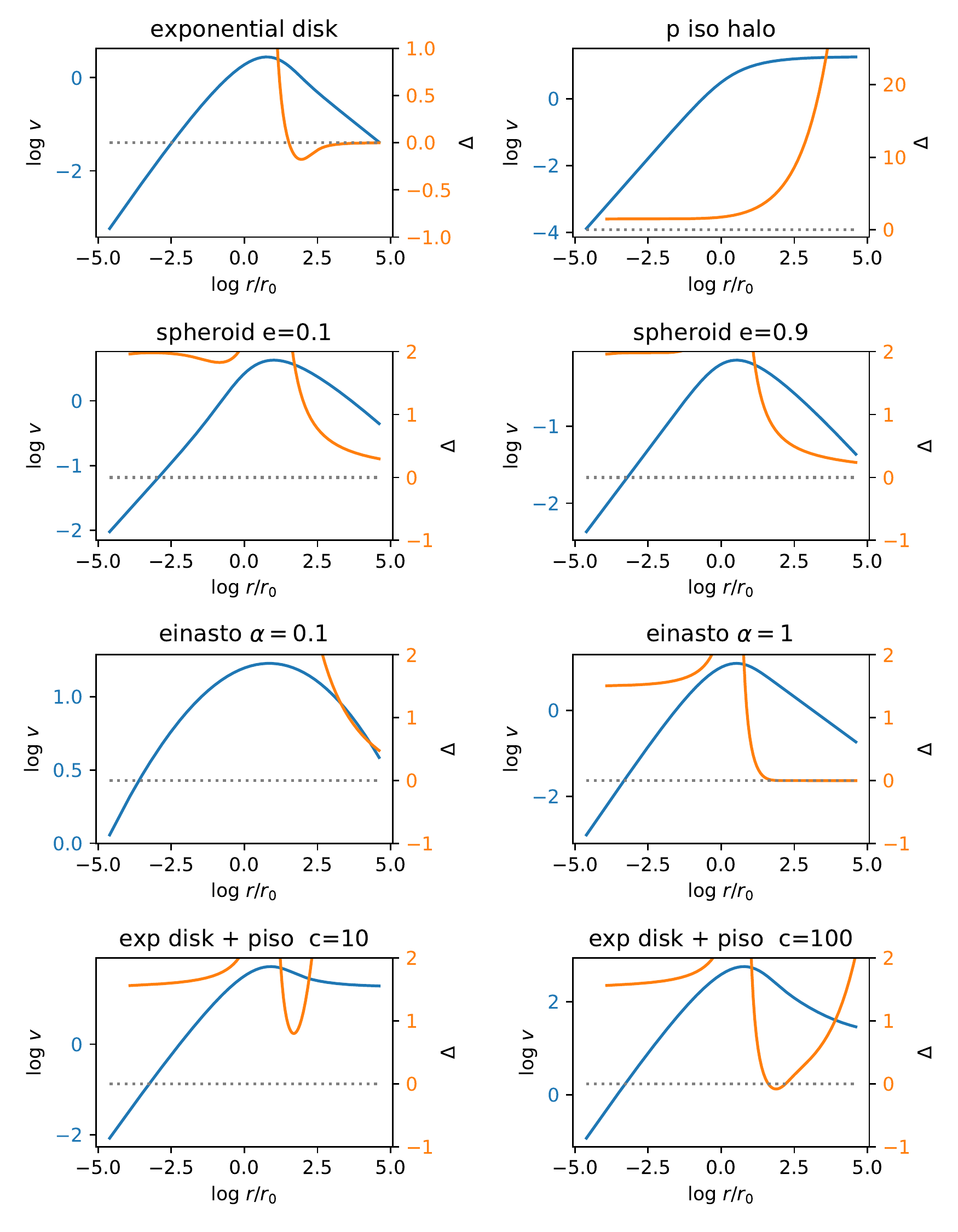}
\caption{Rotation curve and asphericity diagnostic for various mass distributions; Blue: rotation curve, Orange: asphericity diagnostic $\Delta$. Sphericity is violated when $\Delta<0$ and the slope of the log-log rotation curve is less than 1 (eq.\eqref{delta}).  The second row left plot corresponds to an almost spherical spheroid while the right one corresponds to an almost flat spheroid. Note that even thought these spheroids are not spherically symmetric mass distributions, $\Delta$ is positive and this quantity cannot be used to detect asphericity in this case. The two bottom models are a superposition of a pseudo-isothermal halo and an exponential disk with the same scale radius $R_0$ and $c=\frac{\Sigma_0}{\rho_0R_0}$. The asphericity is detectable via a negative $\Delta$ iff $c>52.55$.  }
\label{fig:examples}
\end{figure}

In Figures \ref{fig:annulus} and \ref{fig:examples} we show the asphericity diagnostic $\Delta$ for several different mass models: exponential disk, pseudo-isothermal halo, Hubble oblate spheroid, Einasto halo and a superposition of an exponential disk with the pseudo-isothermal (piso) halo (Table \ref{tab:models}). The exponential disk violates significantly the spherical inequality. The Hubble oblate spheroid is an example of a non-spherical distribution that doesn't violate the spherical inequality. When a halo is added to an exponential disk, if the halo is massive enough, the asphericity becomes undetectable. We illustrate this phenomena with a piso halo. If the exponential disk has the same scale radius $r_0$ as the piso halo, the asphericity is detectable iff $c=\frac{\Sigma_0}{\rho_0r_0}>52.55$.

\begin{center}
\begin{table}
	\begin{tabular}{ | m{4cm}  | m{6cm}  | m{4cm}  | }
	\hline
    	Mass model & Radial density & Comments \\\hline 
		Exponential disk & $\Sigma(r) = \Sigma_0 e^{-r/r_0} $& detectable sphericity violation  \\\hline
		Pseudo isothemal halo & $\rho(r) = \frac{\rho_0}{(1+r/r_0)^2}$ & spherical \\\hline 
		Oblate spheroid (Hubble profile) \citep{Binney1987} &$\rho(m^2)=\rho_0\left(1+\left(\frac{m}{r_0}\right)^2\right)^{-3/2}$ with $m^2=R^2+\frac{z^2}{1-e^2}$&$e=1$:flat, $e=0$:spherical; no detectable sphericity violation  \\\hline
		Einasto halo \citep{Einasto1965}& $\rho(r) = \rho_0 \exp\left(-\frac{2}{\alpha} \left(\frac{r}{r_0}\right)^\alpha-1 \right)$ & spherical \\\hline
		Exponential disk + piso halo & $\rho(r,z)=\rho_{piso}(r)+\delta(z)\Sigma_{exp}(r)$ & superposition of a disk and a halo 
        \\\hline
		Annulus & $\Sigma(r) = \Sigma_0 \text{rect}(\frac{r-r_0}{d}) $ \text{with  } $\text{rect}(r)=\begin{cases} 
      1 & |x|\leq \frac{1}{2} \\
      0 & \text{otherwise}\end{cases}$  &  produces a negative inner rotation curve\\
	\hline
	\end{tabular}
\caption{\label{tab:models}Mass distribution examples. $r$ is the radial (spherical) coordinate, $R$ is the radial distance in cylindrical coordinates and $z$ is the vertical distance.}

\end{table}

\end{center}

\newgeometry{left=1cm,bottom=2cm,top=1cm,right=1cm}

\newgeometry{left=1cm,bottom=2cm,top=1cm,right=1cm}
\restoregeometry

\section{Asphericity of the DM distribution in the SPARC galaxies}
\label{sec:sparc}
\label{sparc}
In this section, we compute $\Delta_{ij}$ \eqref{delta_discrete} for the DM contribution of the SPARC galaxies in order to seek unambigiously non-spherical DM structures.
The contribution of the DM to the rotation curve is
\begin{equation}
    v_\textup{DM}^2 = v_\textup{obs}^2-\Upsilon_\textup{disk} v_\textup{disk}^2-\Upsilon_\textup{bulge} v_\textup{bulge}^2-v_\textup{gas}^2
    \label{eq:vdm}
\end{equation} 

where $v_\textup{obs}$ is the observed circular velocity, $v_\textup{disk}$ and $v_\textup{bulge}$ are the contributions of the stars to the observed velocity assuming a stellar mass-to-light ratio of 1$M_\odot/L_\odot$ and $v_\textup{gas}$ is the contribution of the gas. $\Upsilon_\textup{disk}$ and $\Upsilon_\textup{bulge}$ are the stellar disk and bulge mass to light ratios. Note that $v_i$ is not the velocity of the $i$th component. It is the contribution of the $i$th component to the total circular velocity and is derived by computing the gravitational potential of the $i$th component from the mass densities. The total gravitational potential is the sum of the contributions of each mass component and $v_{obs}^2 = \sum_{i} v_i^2$.
The main unknown is $\Upsilon$. In this work, $\Upsilon$ is estimated by individually fitting each rotation curve with an Einasto halo \citep{Einasto1965}. We use the Einasto model because it can reproduce a great diversity of spherical halo shapes and has been show to fit the SPARC data particularly well \citep{Li2020, Loizeau2021}. In order to obtain a realistic mean mass to light ratio around $0.5$ as suggested by stellar population synthesis \citep{Schombert2014, McGaugh2014, Meidt2014}, the following term is added to the $\chi^2$ when fitting, as in \cite{Loizeau2021}:
\begin{equation}
    \chi^2_\Upsilon = \left(\frac{\Upsilon_{*, i}-\overline\Upsilon_*}{\sigma_{\Upsilon_*}}\right)^2
    \label{eq:chi2upsilon}
\end{equation} 
where $\overline\Upsilon_*=0.5$ is the assumed mean mass to light ratio and $\sigma_{\Upsilon_*}$ is the standard deviation of $\overline\Upsilon_*$ values in an ensemble of galaxies, estimated to be $\sigma_{\Upsilon_*}=0.125$; \footnote{In equation \eqref{eq:chi2upsilon}, $\sigma_{\Upsilon_*}$ is a global parameter that represents the variance in the true stellar mass to light ratio of the ensemble of galaxies. Recall that each galaxy's stellar mass to light ratio is determined by fitting a Einasto halo to the rotation curve so $\sigma_{\Upsilon_*}$ is not the error on $\Upsilon_*$ for each individual fits.
Since this error is not straightforward to estimate and the overall asphericity diagnostic significance is not sensitive to $\overline\Upsilon_*$, we do not include this error in the estimate of the error on $\Delta_{ij}$ that is then fully determined by the error on the rotation curve measurements (equation \eqref{eq:discrete_error}).} $\Upsilon_{*,i}=\Upsilon_{\textup{disk},i}$ is the disk stellar mass to light ratio of the $i$th galaxy and $\Upsilon_{\textup{bulge},i}=1.4\Upsilon_{*, i}$ is the ratio for the bulge; see \citet{Loizeau2021} for more information about these choices.  The Einasto fits have 4 free parameters: 3 halo parameters and the stellar mass to light ratio.

\begin{figure}[H]
\centering
\includegraphics[width=0.9\textwidth]{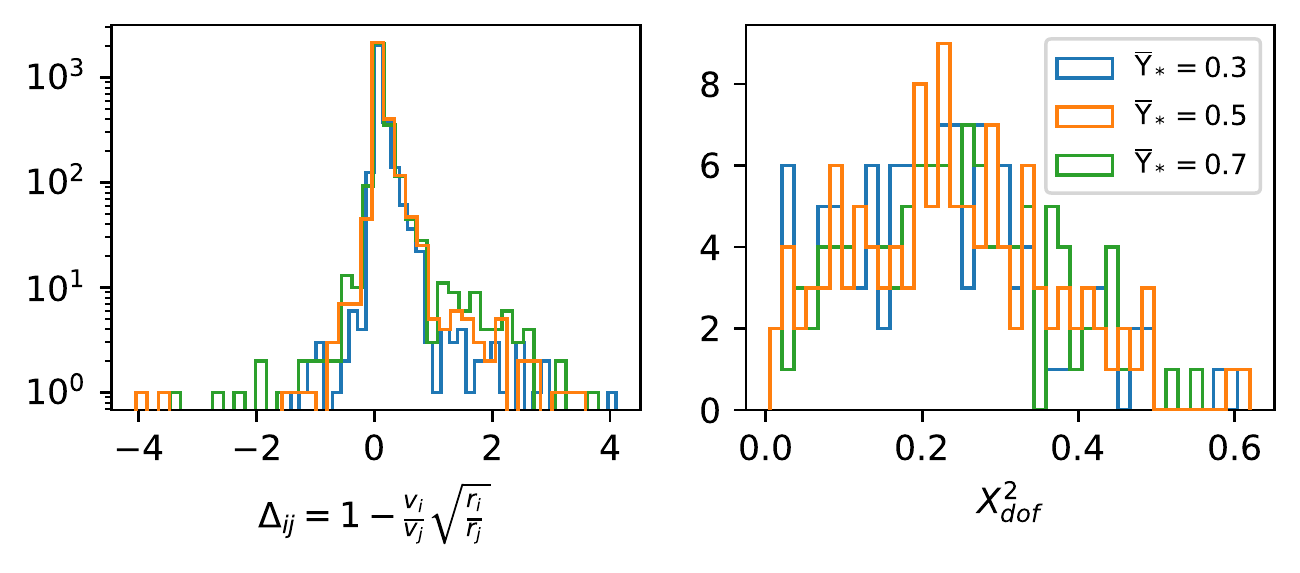}
\caption{\textit{Left}: Distribution of the $\Delta_{ij}(v_{DM})$ values for all $i,j$ neighbors datapoints of the SPARC rotation curves.  \textit{Right}: Distribution of $\chi^2_{dof}$ per galaxy for three values of mean stellar mass to light ratios, showing that the choice of mean $\Upsilon^*$ does not significantly change the conclusions. The individual mass to light ratios are determined using Einasto fits.}
\label{fig:delta_distrib}
\end{figure}

For each galaxy we compute a set of $\Delta_{ij}$ where $i,j$ are neighbor data points, then we combine the significances of these values in order to determine the significance that a given galaxy violates sphericity.
The error on the DM contribution that is used in eq \eqref{eq:discrete_error} is
\begin{equation}
    \delta v_{DM} = \frac{v_{obs}}{v_{DM}}\delta v_{obs}.
\end{equation}
The p-value that one data-point $k$ does not violate sphericity (i.e that $\Delta$ is positive) is the Gaussian integral
\begin{align}
    p_k&=\frac{1}{\sigma_k\sqrt{2\pi}} \int_0^\infty \exp{\left(-\frac{1}{2}\left(\frac{x-\Delta_k}{\sigma_k}\right)^2\right)}dx\\
    &= \frac{1}{2}\left(1+\text{Erf}\left(\frac{\Delta_k}{\sqrt 2 \sigma_k}\right) \right)
\end{align}
with $\Delta_k=\Delta_{k,k+1}$ and $\sigma_k=\delta\Delta_k$ (defined in eq. \eqref{eq:discrete_error}). Note that $p_k\sim0$ means that data-point $k$ shows significant sphericity violation while $p_k\sim1$ means that data-point $k$ does not show sphericity violation.
Using Fisher's method for each galaxy, define
\begin{equation}
    \chi^2= -2\sum_{k=1}^n \ln(p_k).
    \label{eq:chi2}
\end{equation}
If the $p_k$ are independent random variables drawn from the uniform distribution $[0,1]$, then $\chi^2$ follows a $\chi^2$ distribution with $2n$ degrees of freedom. Hence we define $\chi^2_{dof}=\frac{1}{2n}\chi^2$. Now we see that $\chi^2_{dof}\sim 1$ implies that there is no preference for asphericity nor sphericity, while $\chi^2_{dof}>1$ implies that the $p_k$ are biased toward 0 and that there is a significant preference for asphericity.
We find that $\chi^2_{dof} < 1$ for every galaxy in the SPARC dataset and sphericity is not manifestly violated within the measurement uncertainties;  this is shown in figure \ref{fig:delta_distrib}. In figures \ref{fig:sample1} and \ref{fig:sample2}, we show a sample of 16 galaxies from SPARC that exhibit at least one significantly negative $\Delta$ value.

Given the error provided by SPARC, these negative $\Delta$ values are not sufficient to give evidence for asphericity. More precise measurements of these particular galaxies' rotation curves are needed in order to determine whether any of these galaxies may contain non-spherical DM structures of the types detectable through a region of negative $\Delta$.

Given the present large error bars on the $\Delta$ values from SPARC, the statistical indicator derived above is not very useful for identifying promising candidate galaxies which may have DM structure detectable through a negative $\Delta$.
As an alternate approach, we use our fit to the SPARC rotation curves for the hadronically interacting dark matter (HIDM) model developed in \cite{Loizeau2021}, and calculate $\Delta(r)$ for the model DM distribution. This model includes a piso halo and a DM disk that scales the interaction rate per unit volume between  the gas particles in the gas disk and DM particles in the DM halo. The radial density of the DM disk is 
\begin{equation}
\Sigma_{\rm DM}(r) =\zeta \, \rho_{\textup{halo}}(r)\, v_{\textup{obs}}(r)~\Sigma_{\rm gas}(r)
\label{eq:sigmadm}
\end{equation}
where $\zeta$ is a galaxy-dependent free parameter.
This leads to a model where the total observed velocity is 
\begin{equation}
v_\textup{HIDM}^2 = \Upsilon_\textup{disk} v_\textup{disk}^2+\Upsilon_\textup{bulge}v_\textup{bulge}^2+v_\textup{gas}^2+v_\textup{DMdisk}^2+v_\textup{pIso}^2
\end{equation}
where $v_\textup{pIso}=4\pi G \rho_{0} R_{c}^2\Big(1-\frac{R_{c}}{r} tan^{-1}\big(\frac{r}{R_{c}}\big)\Big)$ with free parameters $R_{c}$ and $\rho_{0}$; the contribution of the DM disk $v_\textup{DMdisk}$ is derived from equation \eqref{eq:sigmadm}. The model has a total of 4 free parameters $R_{c},\rho_{0}$, $\zeta$ and $\Upsilon_*$ per galaxy. The fitting procedure is explained in more detail in \cite{Loizeau2021}.

\begin{figure}[H]
\centering
\includegraphics[width=1\textwidth]{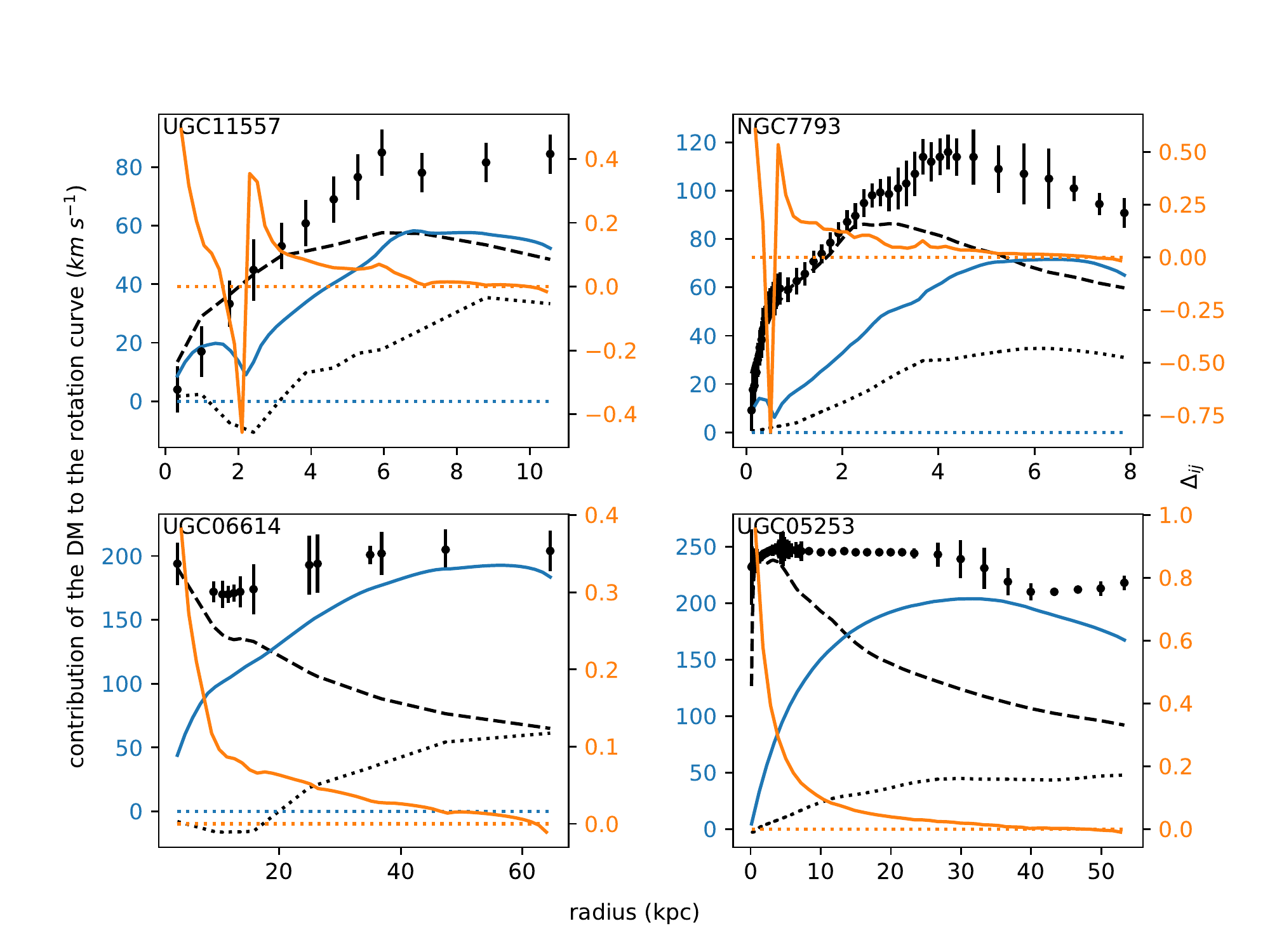}
\caption{Sample of 4 galaxies that have a particularly massive DM disk in the HIDM-IS model. Black markers: observed rotation curve; dashes: stellar contribution displayed with the optimal $\Upsilon_*$; dots: gas contribution.  (Since stars form from gas, it is natural that in UGC11557 one sees similar structure in stars as in gas.) The blue line is the model DM contribution that is the superposition of a disk and a halo. The orange line is the asphericity diagnostic derived from this DM rotation curve. 
The presence of a non-spherical DM distribution can in principle be detectable in the top two, using the asymmetry diagnostic introduced here, but the signature would not be strong enough in the lower pair.  
}
\label{fig:hidm}
\end{figure}

\begin{figure}
\centering
\includegraphics[width=0.9\textwidth]{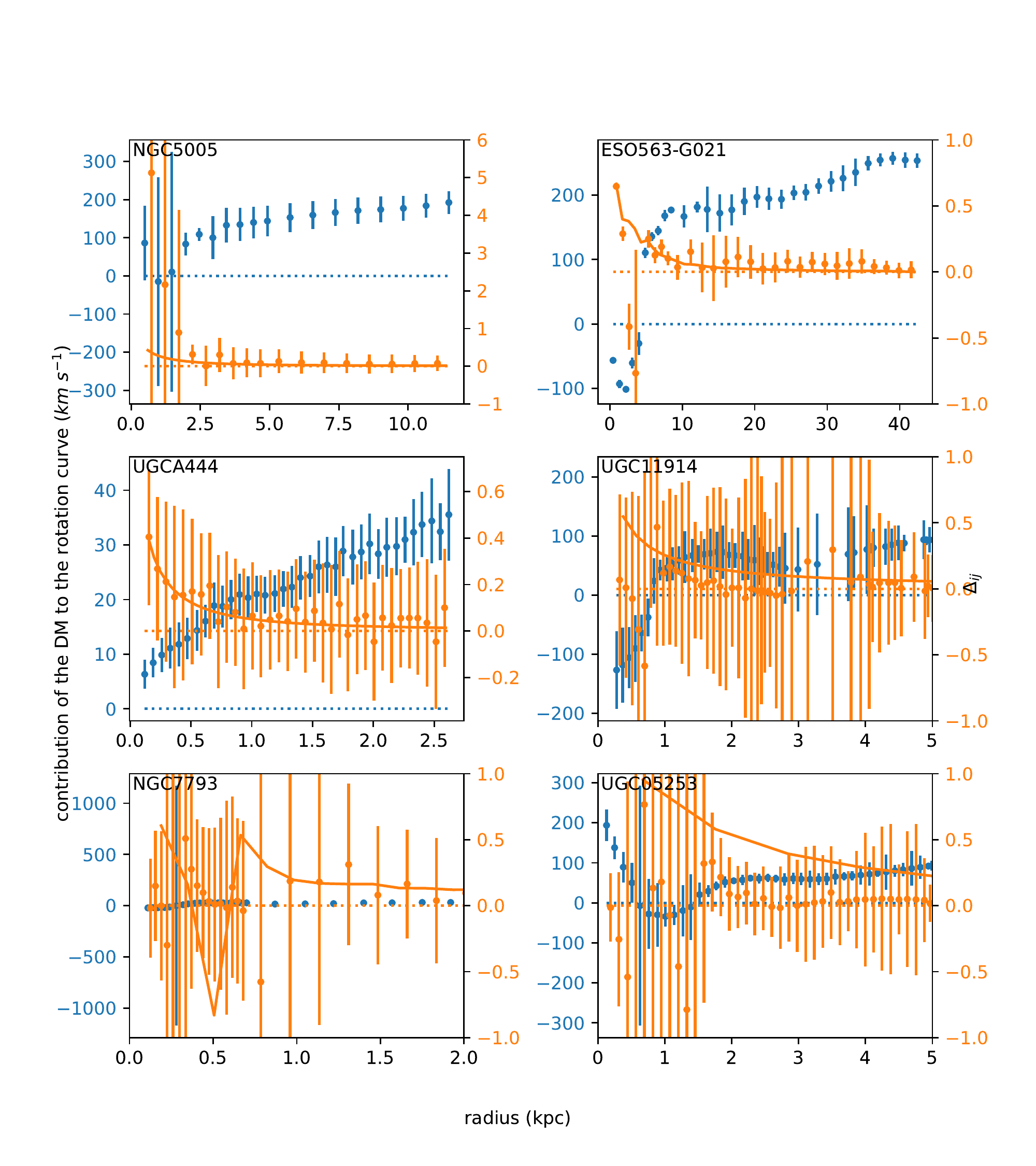}
\caption{Sample of 6 galaxies with the highest likelihood of showing a non-trivial asphericity diagnostic with better data, based on their current $\chi^2_{dof}$ values (eq. \eqref{eq:chi2}) based on the SPARC measurements. Blue markers: contribution of the DM to the rotation curve. Orange markers: discrete asphericity diagnostic $\Delta_{ij}$. Orange solid line: asphericity diagnostic predicted by the HIDM model fit. A significantly negative $\Delta_{ij}$ would show the presence of DM structures like rings, or disks that make the rotation curve drop faster than a spherical distribution. The error is computed assuming that there is no error on the mass to light ratio determination using equation \eqref{eq:discrete_error}. Different mass to light ratios imply different DM contributions and therefore different $\Delta_{ij}$, so in a future attempt to detect asphericity, the mass-to-light ratio should be determined by stellar population synthesis modeling for each galaxy to be studied.    }
\label{fig:sample1}
\end{figure}

\begin{figure}
\centering
\includegraphics[width=0.9\textwidth]{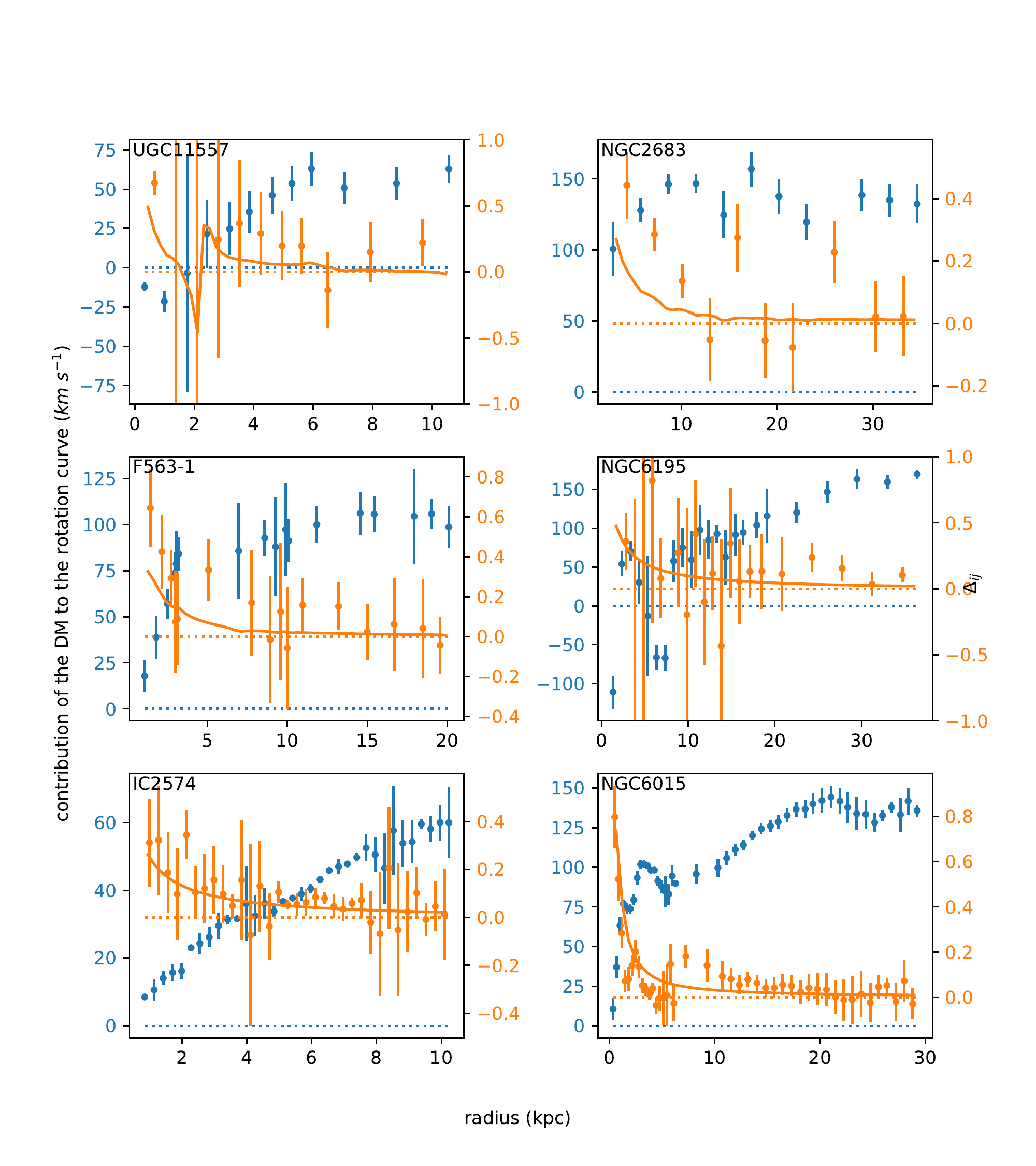}
\caption{Sample of 6 galaxies with the highest percentage of negative $\Delta$ values. Same legend as figure \ref{fig:sample1}. }
\label{fig:sample2}
\end{figure}

These fits allows us to select 4 galaxies that have particularly massive DM disks (Figure \ref{fig:hidm}). 
The two galaxies in the upper row should have an in-principle-detectable-negative asphericity diagnostic stemming from their non-spherical DM component, according to our HIDM model fit. (While the fit to UGC06614 yields a small negative value of $\Delta$ at large radius, it may be undetectably small.) This means that if the model holds and if the uncertainty on the measurements were low enough, it would be possible to detect non-spherical structures by measuring the asphericity diagnostic for their rotation curves.  Fit results for a selection of other galaxies are summarized in table \ref{tab:data}, showing that a number of other galaxies are also good candidates to display a negative $\Delta$.

\section{Conclusion}
\label{sec:conclusion}
We have introduced an asphericity diagnostic $\Delta$, whose positivity is a necessary condition for a spherically symmetric distribution. In section \ref{examples} we showed that the asphericity of some mass distributions such as exponential disks and rings can be detected using this tool, via a negative value of asphericity parameter $\Delta$.  

In section \ref{sparc}, the diagnostic is tested on real rotation curves in the SPARC dataset in order to look for non-spherical dark matter structures. Overall, there is no strong evidence for non-spherical DM structures in the studied dataset given the precision of the measurements. However, with more precise measurements this tool could be a useful diagnostic of non-spherical DM structures that could emerge from baryon interactions. Given that the distribution of significance $\chi^2_{dof}$ of the asphericity diagnostic peaks around $0.2$ (figure \ref{fig:delta_distrib} right), the exact same measurements with an error $\sqrt 5$ smaller could show significant asphericity. Note that this is only a lower bound on the required increase in needed precision assuming identical measurement values.

The five SPARC galaxies with the highest dark matter asphericity significance are NGC5005, UGCA444, UGC11914, NGC7793 and UGC05253, with $\chi^2_{dof}=0.48, 0.49, 0.61, 0.59, 0.48$ respectively; $\chi^2_{dof}$ is defined (see Sec.~\ref{sparc}) such that $\chi^2_{dof} \geq 1$ would be a robust indicator of asphericity. The five galaxies with the highest percentage of negative $\Delta$ values are NGC2683, F563-1, UGC11914, NGC6015, NGC7793 with $30,18,21,20,33\%$ of negative values respectively.  Currently, the error bars on the rotation curves are too high for these to be significant evidence of non-spherically symmetric dark matter.  Moreover based on the HIDM fit of \cite{Loizeau2021} the DM distribution in these galaxies -- in spite of not being spherical -- would not be expected to be manifestly asymmetric in the manner needed for the asphericity diagnostic to reveal it.  More results for individual galaxies are summarized in table \ref{tab:data}.

Using the HIDM model fit from \cite{Loizeau2021}, we identified four galaxies that are compatible with a particularly heavy dark mater disk built up through DM accretion in the baryonic disk: UGC11557, NGC7793, UGC06614 and UGC05253.  These are displayed in Fig. 4 where it can be seen that NGC7793 and UGC05253 may be good candidates for future detection of small-scale structure in the dark matter distribution using the asphericity diagnostic, if measurement accuracy can be sufficiently improved.  

Independently of the HIDM model of \cite{Loizeau2021}, galaxies displaying a strong variation in the radial distribution of gas are the best candidates for having $\Delta < 0$, if any galaxies do.  A DM ring would a smoking gun for HIDM interactions, and would be strongest in galaxies with the strongest gas rings. The measurement error bars in Fig. 4 include rough systematic errors assigned by the SPARC collaboration which can be reduced by more precise modeling of the radial velocity data, suggesting that the required reduction of errors in $\Delta$ can be within reach.

In sum, to most effectively investigate whether there is a non-trivial small-scale structure in DM, as a result of DM collecting with gas, the following observational procedure is suggested:
\begin{enumerate}
\item Identify galaxies for which good rotation curves can be obtained (viewing angle neither too edge-on or flat-on).
\item Identify a subset of galaxies with strong radial variations in gas or gas+stars, especially ring-like structures.
\item Make accurate measurements of the circular velocity as well as the mass distribution of gas and stars in the radial region enclosing the gas structure, with high radial resolution.  It is crucial to measure all of these accurately, to accurately find the DM contribution.  It is not necessary to improve the measurements at large radii far beyond the gas, which is fortunate since tracing the rotation curve at large radius can be difficult.
\item Perform detailed stellar-population modeling of these individual galaxies, in order to accurately assign the correct stellar mass-to-light ratio, which may change as a function of radius.
\end{enumerate}
Using this data, the DM contribution to the rotation curves of the selected galaxies can be determined, so the local asphericity diagnostic $\Delta$ introduced
here can be measured.  Finding that $\Delta < 0$ with sufficiently small statistical and systematic error bars, even at one radial position, would provide unambiguous proof that DM has aspherical structure. (Of course, observing it at a sequence of radii and in multiple galaxies would be needed to become convinced.) The last two columns of Table~\ref{tab:data} can be helpful in selecting promising galaxies for detecting a negative $\Delta$, based on the HIDM modeling of \cite{Loizeau2021}.  Perhaps the most favorable three in the SPARC dataset, in terms of a strong signal covering a significant radial region according to the HIDM fit, would be NGC2915, 6195 and 3521, while independently of the model, NGC7793 and UGC1155 and good candidates.

\subsection*{Acknowledgments}
The research of G.R.F. is supported in part by NSF-PHY-2013199 and by the Simons Foundation.

\newgeometry{left=1cm,bottom=2cm,top=1cm,right=1cm}

\restoregeometry

\begin{sidewaystable}
\begin{tabular}{ l  l | l  l | l  l | l  l | l  l } 
name & $\chi^2_{dof}$ & name & neg $\Delta$ & name & min $\Delta$ & name & neg $\Delta$ & name & min $\Delta$  \\ 
 &  &  &  &  &  &  & HIDM&  &  HIDM  \\ 
\hline 
UGC11914&0.620&NGC7793&0.333&NGC7793&-54.151&NGC7793&0.333&NGC4088&-14.914\\ 
NGC7793&0.598&NGC2683&0.300&UGC02885&-4.064&NGC3521&0.325&NGC2915&-4.485\\ 
UGCA444&0.494&UGC11914&0.219&UGC11557&-3.589&NGC2683&0.300&NGC6195&-3.999\\ 
NGC5005&0.484&NGC6015&0.209&NGC5585&-1.517&NGC2955&0.261&UGC05253&-2.994\\ 
UGC05253&0.482&F563-1&0.188&NGC4088&-1.338&NGC6195&0.227&NGC3521&-2.165\\ 
ESO563-G021&0.467&NGC6195&0.182&NGC6195&-1.043&UGC11914&0.219&NGC0891&-2.112\\ 
NGC2915&0.461&IC2574&0.182&ESO563-G021&-0.885&F563-1&0.188&UGC06787&-0.964\\ 
NGC2683&0.454&UGC11557&0.182&UGC05253&-0.788&IC2574&0.182&UGC02885&-0.952\\ 
NGC3521&0.449&NGC3726&0.182&NGC6946&-0.704&UGC11557&0.182&NGC5585&-0.896\\ 
NGC6946&0.433&NGC6946&0.175&UGC11914&-0.584&NGC2915&0.172&NGC7793&-0.883\\ 
NGC4013&0.426&UGC11455&0.143&UGC03546&-0.560&NGC4013&0.171&NGC7331&-0.788\\ 
NGC4559&0.415&NGC5033&0.143&UGC02953&-0.485&NGC2998&0.167&UGC08699&-0.707\\ 
UGC02953&0.409&UGC05253&0.139&F583-1&-0.321&NGC5033&0.143&UGC11914&-0.585\\ 
NGC6195&0.406&NGC2915&0.138&UGC08699&-0.303&UGC02916&0.143&UGC02953&-0.520\\ 
UGC12506&0.405&UGC03546&0.138&NGC0289&-0.302&NGC6946&0.140&UGC03546&-0.473\\ 
NGC0289&0.391&UGC12506&0.133&NGC4157&-0.211&NGC6015&0.140&NGC2955&-0.421\\ 
F563-1&0.384&NGC2955&0.130&NGC4013&-0.106&UGC12506&0.133&UGC11557&-0.390\\ 
NGC2976&0.376&NGC2976&0.115&NGC3726&-0.103&UGC02953&0.132&UGC02916&-0.345\\ 
UGC11557&0.374&NGC4013&0.114&UGC11455&-0.083&NGC2403&0.125&F583-1&-0.327\\ 
F583-1&0.369&UGC02953&0.114&UGC09037&-0.080&UGC05253&0.125&NGC2998&-0.298\\ 
\end{tabular}
\caption{\label{tab:data} List of galaxies with significant non-sperical DM component. Each column is sorted according to one indicator and only the 20 most significant galaxies are displayed. $\chi^2_{dof}$ is the asphericity diagnostic defined in equation $\eqref{eq:chi2}$. 'neg $\Delta$' is the fraction of negative $\Delta$ values in the rotation curve. 'min $\Delta$' is the minimum $\Delta$ values in the rotation curve. HIDM indicates that the indicator is calculated based on a HIDM model fit to the rotation curve as explained in the end of section \ref{sec:sparc}. Note that the full dataset has 121 galaxies.}
\end{sidewaystable}

\printbibliography
\end{document}